\documentclass[a4,11pt]{cipro1}
\usepackage{amsmath,amsxtra,amssymb,latexsym, amscd,color}
\usepackage[mathscr]{eucal}
\usepackage{graphicx}
\begin{document}
\def\figup{\pagebreak$\quad$\vskip-0.7cm}
\def\Mr{\MakeUppercase}
\def\vsa{\vskip-0.2cm}
\def\vssa{\vskip-0.4cm}
\def\vs{\vskip0.2cm}
\def\vsm{\vskip0.1cm}
\def\vss{\vskip0.4cm}
\def\bce{\begin{center}}
\def\ece{\end{center}}
\def\n{\noindent}
\def\nn{\nonumber}
\def\ce{\centerline}
\def\disp{\displaystyle}
\def\abs{\begin{abstract}\rightskip-1.2cm\it}
\def\eabs{\end{abstract}}
\parindent=1.05cm
\title
{\large THE BACKGROUND FIELD METHOD APPLIED TO COSMOLOGICAL PHASE
TRANSITION}
\maketitle
\markboth{PHAN HONG LIEN, NGUYEN NHU XUAN}{THE BACKGROUND FIELD
METHOD APPLIED TO COSMOLOGY}
\begin{center}
PHAN HONG LIEN, NGUYEN NHU XUAN\\
{\it Le Quy Don Technical University}
\end{center}
\abs   
In this paper, the cosmological phase transition is investigated by
background gauge field method. As a continuation of previous our
work, some numerical results and graphic solutions at $T\neq 0$ are
presented. Hence the mechanism of cosmological phase transition in
the early Universe is considered. It is shown that the breaking of
symmetry significantly depend on the nonzero temperature and
chemical potential. Furthermore, it is the first order of phase
transition. Non - restoration of symmetry in hot gauge theories for
Cosmology.\eabs
\section{INTRODUCTION}
\indent Phase transition is a complicate physical process, its
nature is non pertubative phenomenon [1] - [5]. Therefore, it is
worth to mention that the finite temperature effective action basing
on functional integral as a method which provides a general
approximation beyond one loop and higher free energy density in the
pertubative as well as non-pertubative sector. In particular, it
plays an important role in the investigation of phase transition and
non-equilibrium phenomena [6],[7], as dynamics of processes.\\
\indent Recently there has been considerable interest in the
symmetry in both hot scalar field theories and hot gauge theories
for cosmology [8],[10]. It is shown that the phase transition in the
early Universe could be described by non-Abelian gauge theories at
high temperature [11]. Furthermore, the contribution to the free
energy only by calculated by using non-pertubative methods. However,
the effective potential of gauge theories may fail to be gauge
because it generally does depend on the $\xi$ - gauge. Therefore the
background field method have been applied to compute quantum effects
without losing gauge invariance [12],[13].\\
\indent Our main aim is to apply the background gauge field method
at high temperature to investigation of cosmological phase
transition. In this connection, it is possible to consider this
paper as being complementary numerical computation to result of
previous our work.\\
\indent This paper is organized as follows. In section II, we
present the formalism of effective action and background gauge field
method. The one loop free energy density at nonzero temperature an
chemical potential is obtained in section III. Section VI is devoted
to investigation the cosmological phase transition from some
numerical calculations and graphic solutions. Our conclusion is
summarized in section V.
\section{FORMALISM}
\indent We start from the Lagrangian density
\begin{equation}\label{e21}
    \begin{split}
        L_0=&-\frac{1}{4}F_{\mu\nu}^aF^{a\mu\nu}+\mathbf{\bar{\Psi}}(i\gamma^\mu D_\mu
      - G_i\Phi_i)\mathbf{\Psi}\\
      &+\left[(D_\mu-i\mu\delta_{\mu 0})\Phi_i\right]^+\left[(D^\mu-i\mu\delta^{\mu 0})
      \Phi_i\right]-m^2\Phi^+_i\Phi_i-\lambda(\Phi^+_i\Phi_i)^2\\
      &-\frac{1}{2\xi}\left(\partial^\mu A_\mu^a\right)^2-\partial_\mu
      \omega_a^*\partial^\mu\omega_a+f_{abc}(\partial_\mu
      \omega_a^*)A_\mu^b\omega^c.
    \end{split}
\end{equation}
where $\mathbf{\bar{\Psi}}, \mathbf{\Psi}$ are multiplet of fermion
fields, $\Phi_i (i=1,2 \ldots n)$ are components of scalar fields,
$A_\mu$ - gauge fields and $\omega,\omega^*$ - ghost fields. Here
$\mu$ is chemical potential, $G_i$ and $\lambda$ are coupling
constants, $\lambda>0$
\begin{equation*}
    \begin{split}
       D_\mu\equiv&\hspace{3pt}\partial_\mu-iT^aA_\mu^a,\\
       F_{\mu\nu}^a\equiv&\hspace{3pt}\partial_\mu A_\nu^a-\partial_\nu
       A_\mu^a+f_{abc}A_\mu^b A_\nu^c,
     \end{split}
\end{equation*}
where $T_a$ are group generators, $f_{abc}$ are structure constants
which satisfy Lie algebra
\begin{align}
    f_{abc}f_{dbc}=&\hspace{3pt} g^2C_A\delta_{ad} \label{e23a}\\
    Tr(T_aT_b)=&\hspace{3pt}g^2C_F\delta_{ab}\label{e24a}
\end{align}
with $C_A$ is numerical constant of gauge group, $C_A=N$ for
$SU(N)$, $C_F$ is representation of this group.\\
 \indent The fields are shifted by
 \begin{align}
    A_\mu\rightarrow&\hspace{3pt} \mathbf{A}_\mu+A'_\mu; \hspace{1cm}
    \langle0|\mathbf{A}_\mu|0\rangle=const,\hspace{14pt}\langle0|A'_\mu|0\rangle=0,
    \label{e22}\\
    \Phi_i\rightarrow& \hspace{3pt}\mathbf{\Phi}_i+\Phi'_i; \hspace{1cm}\hspace{5pt}
    \langle0|\mathbf{\Phi}_i|0\rangle\hspace{3pt}=\phi_{0i},\hspace{28pt}\langle0|\Phi'_i|0\rangle=0,
    \label{e23}\\
    \Psi\rightarrow&\hspace{3pt} \mathbf{\Psi}+\Psi'; \hspace{1cm}\hspace{9pt}
    \langle0|\mathbf{\Psi}|0\rangle\hspace{4pt}=\langle0|\mathbf{\Psi}'|0\rangle=0,
    \label{e24}\\
     \omega_a\rightarrow&\hspace{3pt} \boldsymbol{\omega}_a+\omega'_a; \hspace{1cm}\hspace{6pt}
    \langle0|\boldsymbol{\omega}_a|0\rangle=\langle0|\omega'_a|0\rangle=0,
    \label{e25}
 \end{align}
 where $\mathbf{A}_\mu,\mathbf{\Phi},\mathbf{\Psi},\boldsymbol{\omega}_a$
 are the background fields, and $A'_\mu,\Phi',\Psi',\omega'_a$ are the quantum fields,
 which are variables of integration in the functional integral. \\
 \indent It is well known, the background field method allows one to fix a gauge, thereby compute quantum effects without losing explicit gauge invariance [14].\\
\indent Now we consider the effective action in background field for
which $\mathbf{A}_\mu^a$ and $\mathbf{\Phi}_i$ are constant, and
$\mathbf{\Psi}=\mathbf{\bar{\Psi}}=\boldsymbol{\omega}=\boldsymbol{\omega}^*=0$.
The effective action is calculated from the part of the action that
is quadratic in quantum fields $A'_\mu,\Phi',\Psi'$ and
$\omega',\omega'^*$ over which one integrated
 \enlargethispage{2cm}
\begin{equation*}\label{e26}
\begin{split}
I_{quad}=&\int dx L_{quad}=\int
dx\left[-\frac{1}{4}\left(\bar{D}_\mu A_\nu^{'a}-\bar{D}_\nu
A_\mu^{'a}\right)^2-\frac{1}{4}F_{\mu\nu}^af_{abc}A_\mu^{'b}A_\nu^{'c}\right]\\
&-\int dx \bar{\Psi'}\left(\gamma^\mu
D_\mu+\mu\gamma^0+M+g_i\phi'_i\right)\Psi'\\
&+\int
dx\left[\frac{1}{2}\left(D_\mu\phi'_iD^\mu\phi'_i-M_{ij}^2\phi_i^{'2}\right)
-\frac{\lambda}{4}\phi_i^{'4}\right]\\
&-\int dx\left[\frac{1}{2\xi}\left(\bar{D}_\mu
A_\nu^{'a}\right)^2+\left(\bar{D}_\mu
\omega_a^{'*}\right)\left(\bar{D}_\mu
\omega'_a\right)\right]\end{split}
\end{equation*}
\begin{equation}\label{e26}
\begin{split}
I_{quad}=&\frac{1}{2}\int dxdy
A_\mu^{'a}(x)\mathfrak{D}_{\mu\nu}^{ab}(x,y)A_\nu^{'b}(y)
-\int dxdy \Psi'(x)\mathfrak{D}_{ab}(x,y)\Psi'(y)\\
&+\frac{1}{2}\int dxdy \phi'_i(x)\mathfrak{D}_{ik}(x,y)\phi'_k(y)
-\int dxdy\omega_a^{'*}(x)\mathfrak{D}_{ab}(x,y)\omega'_b(y)
\end{split}
\end{equation}
\indent By using the Fourier transformation $\mathfrak{D}(k)=\int dx
e^{ik(x-y)}\mathfrak{D}(x-y)$ the matrices in (\ref{e26}) are given
by
\begin{equation}\label{e27}
\begin{split}
  \mathfrak{D}_{\mu\nu}^{ab}(k)=&g_{\mu\nu}\Bigr[\left(-ik_\rho\delta_{ca}+f_{cda}A_\rho^d\right)
  \left(-ik^\rho\delta_{cb}+f_{ceb}A_\rho^e\right)\\
&-\left(-ik_\nu\delta_{ca}+f_{cda}A_\nu^d\right)\left(-ik_\mu\delta_{cb}+f_{ceb}A_\mu^e\right)
+F_{\mu\nu}^cf_{cab}\Bigr]\\
&+g_{\mu\nu}\delta_{ij}\Phi_i(k)\Phi_j(k)T^aT^b+\mbox{$\epsilon$
terms},
\end{split}
\end{equation}
with $F_{\mu\nu}^a=f_{abc}A_\mu^b A_\nu^c$.
\begin{align}
 \mathfrak{D}(k)=&(-ik\hspace{-6pt}/-iT^aA^a_\mu\hspace{-12pt}/\hspace{9pt}
+M+\mu\gamma_0)+\mbox{$\epsilon$ terms},\label{e28}\\
 \mathfrak{D}_{ab}(k)=&\left(-ik_\rho\delta_{ca}+f_{cda}A_\rho^d\right)
\left(-ik^\rho\delta_{cb}+f_{ceb}A^{\rho b}\right)+\mbox{$\epsilon$ terms},\label{e29}\\
\mathfrak{D}_{ij}(k)=&\left(-ik_\mu-i\mu-iT_aA_\mu^a\right)_i
\left(ik_\mu+i\mu+iT_aA_\mu^a\right)_j-m_{ij}^2-\frac{\lambda}{2}\phi_i\phi_j+\mbox{$\epsilon$
terms},\label{e210}
\end{align}
\indent In momentum representation, the effective action takes the
general form
\begin{equation}\label{e211}
    \begin{split}
     &\Gamma_\beta\left[\psi,\bar{\psi},\phi,A_\mu,\omega,\omega^*\right]
     =I\left[\psi,\bar{\psi},\phi,A_\mu,\omega,\omega^*\right]
     -\frac{i}{2}Trln G_{\mu\nu}^{ab}(k)\\
     &+iTrlnS(k)-\frac{i}{2}Trln\Delta_{ij}(k)+iTrlnD_{ab}(k)+\sum_{n=2}^\infty
\mbox{\emph{n loops 1PI}}.
    \end{split}
\end{equation}
Here the action $I\left[\psi,\bar{\psi},\phi,A_\mu,
\omega,\omega^*\right]$ is given in (\ref{e26}) - (\ref{e210}). The
Trace, the logarithm are taken in functional sense, and the free
propagators are given by
\begin{align}
   S^{-1}(k)&=k\hspace{-6pt}/-M-i\epsilon;\hspace{2.5cm} M=g\nu,\label{e212}\\
  \Delta_{ij}^{-1}(k)&=\delta_{ij}k^2-M_{ij}^2 -i\epsilon;\hspace{1.6cm}M_{ij}^2
  =(\mu^2-m^2)\delta_{ij}-\frac{\lambda}{2}\phi_i\phi_j \label{e213},\\
 \left[G_{0ab}(k)\right]_{\mu\nu}^{-1}&=\left(M^2_{ab}-k^2\delta_{ab}\right)\left[\frac{k_\mu
 k_\nu}{k^2}-g_{\mu\nu}\right]+\left[\delta_{ab}\frac{k^2}{\xi}-M^2_{ab}\right]\frac{k_\mu
 k_\nu}{k^2}, \label{e214}\\
D_{0ab}^{-1}(k)&=\delta_{ab}(k^2-i\epsilon),\hspace{2.5cm}M_{ab}=\frac{1}{2}\delta_{ab}g\nu.
\label{e215}
\end{align}

\section{ONE LOOP THERMAL FREE ENERGY DENSITY}
Next we consider the theory at finite temperature by "imagine time"
formalism.  For $\langle
\mathbf{A}_\mu\rangle_\beta=\delta_{0\mu}A_\mu^0,\langle\mathbf{\Phi}\rangle=\phi_0$
the effective potential is defined by
\begin{equation}\label{e31}
    V_\beta=-\frac{\Gamma_\beta}{\beta\int dx}
\end{equation}
where $\beta=T^{-1}$ (we set Boltzmann constant k = 1). It is just
thermal free energy density, which concerns with the phase
transition at $T=T_c$.\\
\indent In $d=4-2\epsilon$ dimension, the divergent integrals can be
regularized by using the "imagine time" formalism, where nonzero
chemical potential $\mu$ is added to the fermionic Matsubara
frequencies, i.e.
 \begin{equation*}
    i\int\frac{d^4k}{(2\pi)^4}f(k)\rightarrow
    T\sum\hspace{-15pt}\int\frac{d^3k}{(2\pi)^3}f(i\omega_n,\vec{k}),
 \end{equation*}
where $i\omega_n=2\pi nT$ for bosons, $\omega_n=(2n+1)\pi T$ for fermions and $i\omega_n\rightarrow i\omega_n+\mu$.\\
\indent From (\ref{e26}) - (\ref{e211}) and (\ref{e31}) we arrived
at the expression for the effective potential
\begin{equation}\label{e32}
 \begin{split}
   V_\beta=&V_{cl}-\sum_k\hspace{-15pt}\int \left[ln(k^2+M^2)-ln(k^2+M_{ij}^2)-ln(k^2+M_{ab}^2)
   \right]\\
   &+ig^2\left(\frac{11}{12}N-\frac{1}{6}N_F+\frac{1}{12}N_B\right)\sum_k\hspace{-15pt}\int\frac{1}{(k^2)^2}\int dx F_{\mu\nu}^a
   F^{a\mu\nu}\\
   &+\frac{i}{8}g^2\sum_k\hspace{-15pt}\int \frac{1}{(k^2+m^2)^2}\\
   &+\frac{i}{2}g_i\sum_k\hspace{-15pt}\int\sum_p\hspace{-15pt}\int\frac{1}{(p^2+M_1^2)(k^2+M_2^2)
\left[(k+p)^2+M_{ab}^2\right]}.
 \end{split}
\end{equation}
\indent Finally, the one loop thermal free energy density is given
\begin{align}\label{e33}
  V_\beta=&-\frac{1}{2}(\mu^2-m^2)\phi^2+\frac{\lambda}{4}\phi^4-\frac{1}{2}
      \delta_{ab}M_{ab}^2A_{0\mu}^2-\frac{\pi^2T^4}{90}\left(N_B+\frac{7}{8}N_F\right)\notag \\
     &+\frac{T^2}{24}\left\{(\mu^2-m^2-\frac{\lambda}{2}\phi^2)+3TrM_{ab}^2+\frac{1}{2}
     Tr\left[\gamma_0(M+\mu\gamma_0)\gamma_0(M+\mu\gamma_0)\right]\right\}\notag\\
     &-\frac{T}{12\pi}\left(\mathfrak{M}^3+\delta_{ab}M_{ab}^3\right)-\frac{g^2T^3}{48\times 4\pi}
     \left(\mathfrak{M}+\delta_{ab}M_{ab}+2M\right)\\
     &+\frac{g^2}{(4\pi)^2}\left(\frac{11}{12}N-\frac{1}{6}N_F+\frac{1}{12}N_B\right)
     \left(\frac{1}{\epsilon}-2ln\frac{\bar{\eta}}{4\pi
     T}+2\gamma_E\right)\int dx F_{\mu\nu}^aF^{a\mu\nu}.\notag
\end{align}
where $\mathfrak{M}, M_{ab}, M$ are thermal masses, e.g the squared
scalar mass is
\begin{equation}\label{e34}
    \mathfrak{M}^2=(\mu^2-m^2)+\frac{\lambda}{24}T^2-\frac{\lambda}{2}\phi^2.
\end{equation}
The renormalized coupling is given by

\begin{equation}\label{e35}
g_R=g\left[1+\frac{g^2}{4\pi^2}\left(\frac{11}{12}N-\frac{1}{6}N_F+\frac{1}{12}N_B\right)
        \left(\frac{1}{\epsilon}+2ln\frac{\bar{\eta}}{4\pi T}+2\gamma_E\right)\right]+0(g^4)
\end{equation}
i.e the physical coupling $g_R$ increases due to quantum corrections
in the non - Abelian theory. \pagebreak
\section{SOME NUMERICAL RESULTS AT NONZERO TEMPERATURE AND CHEMICAL POTENTIAL}
Let us consider the minimum of the effective potential from which we
can determine the breaking or the restoration of symmetry. \\
\indent In the case of $\mu\leq m$, the free energy density is
minimum at $\Phi_0=0$ at any temperature T, i.e the symmetry is
not broken (Fig. 1).\\
 \indent In Fig. 2 it is easily seen that the high temperature
and nonzero chemical potential significantly affect on the
spontaneous breaking of symmetry. The importance is that it leads to
the cosmological phase transition.\\
\indent In Fig. 3, we can see that the effective squared scalar mass
is change from negative to positive value when the temperature is
high enough. It is due to the contribution of the
$\mathfrak{M}^2T^2$ term in the effective potential. This just is
phenomena of breaking
of symmetry, when it occurs the cosmological phase transition is manifested.\\
\indent In Fig. 4, the free energy is represented as a continuous
function of $T$ and $\mu$ when $\mu\leq m$, but in Fig. 5, it has a
discontinuity when $\mu\geq m$. It is shown by the discontinuous
buffer between the symmetry and its breaking part. This is just the
cosmological phase transition. Furthermore, the restoration of symmetry does appear that means after symmetry was spontaneously broken, the Universe is asymmetry\\
\indent In Fig.6, the effective squared scalar mass depend on the
chemical potential $\mu$ at fixed temperature, which expresses
\textbf{the first order phase transition}.\\
 \vspace{-4cm}
\begin{center}
\begin{tabular}{cc}
 \includegraphics[width=6cm,height=4.5cm]{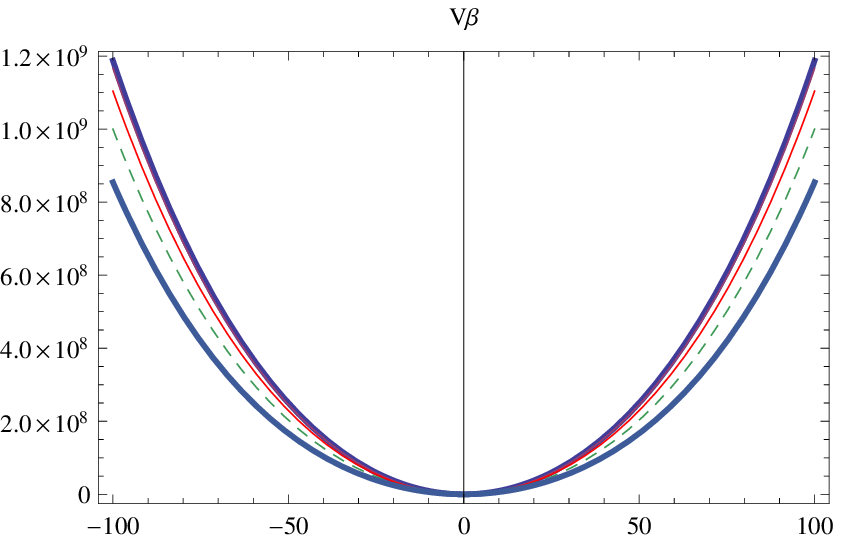}
 &\hspace{1cm}\includegraphics[width=6cm,height=4.5cm]{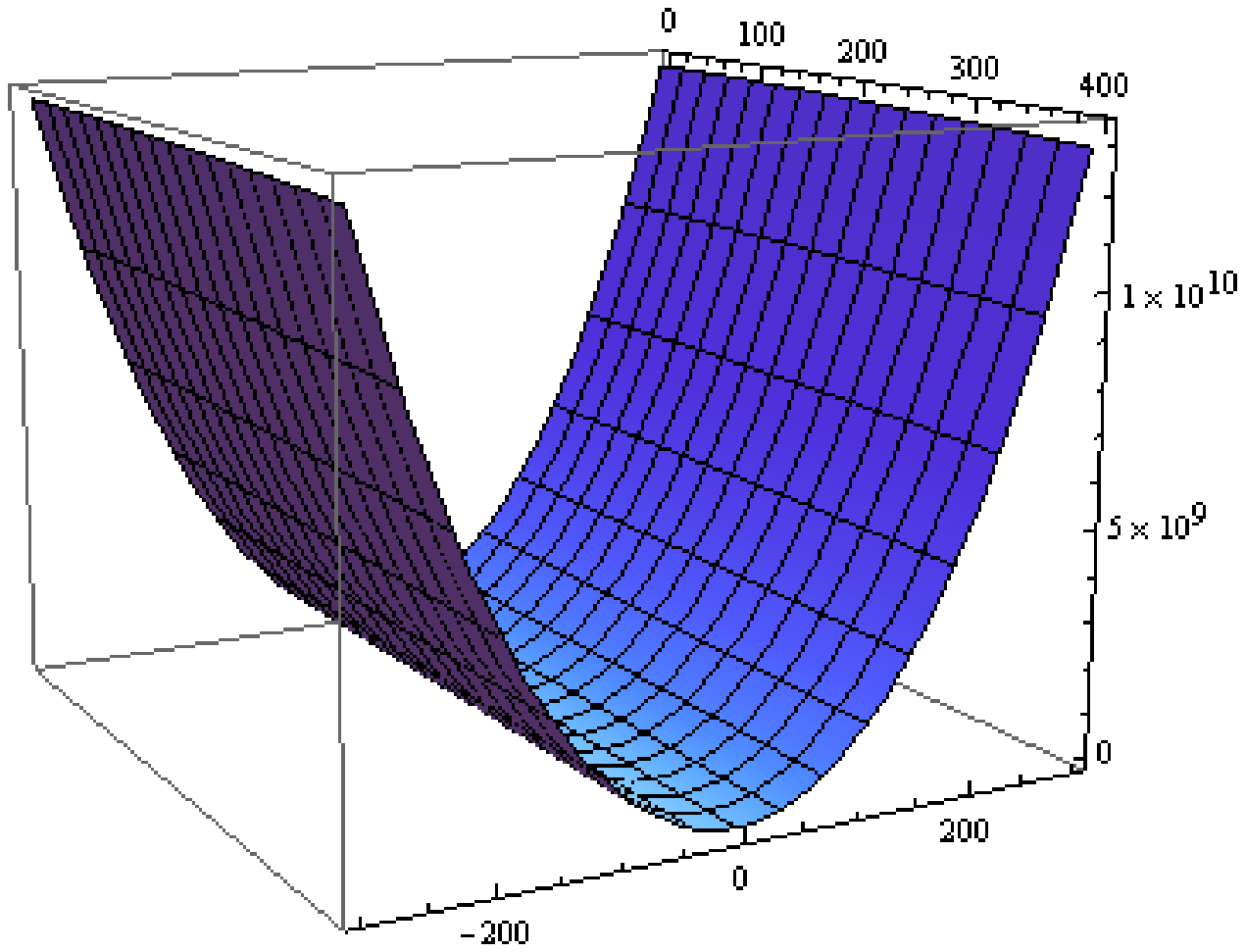}\\
 a)& \hspace{1cm} b)
\end{tabular}
\footnotesize{\textbf{Fig.1.} The effective potential $V_\beta$ as a
function of T and $\Phi$ in the case $\mu/m= 0.5; 1.0$.}\\
\indent a) \footnotesize{$\Phi = -100 \div 100 MeV$ with $T = 0, 100, 200, 400 MeV$ respectively}\\
 \indent b) \footnotesize{$T = 0\div 400$ MeV and $\Phi = -300 \div 300$ MeV}
 \end{center}
 \vspace{-1cm}
 \pagebreak

 \begin{center}
\begin{tabular}{cc}
 \includegraphics[width=6cm,height=4.5cm]{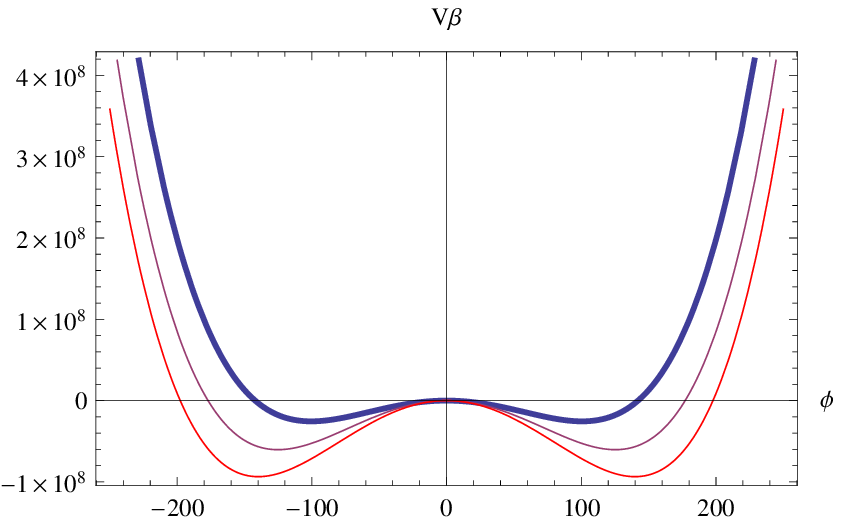}
 &\hspace{1cm}\includegraphics[width=6cm,height=4.5cm]{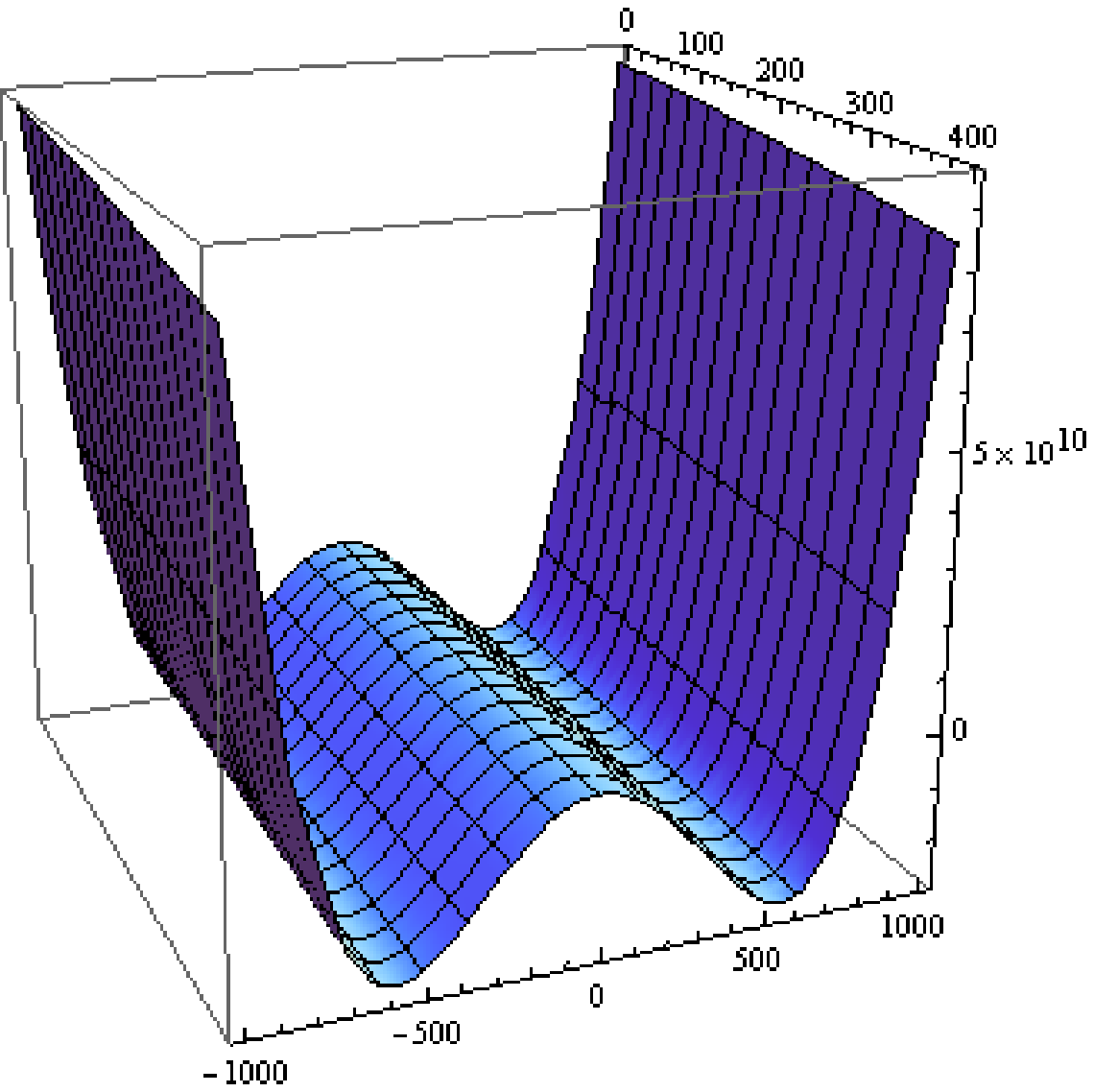}\\
 a)& \hspace{1cm}b)
\end{tabular}
 \footnotesize{\textbf{Fig.2.} The effective potential $V_\beta$ as a
function of T and $\Phi$ in the case $\mu/m= 1.5$.}\\
 \indent \hspace{28pt} a)
\footnotesize{$\Phi = -100 \div 100 MeV$ with $T = 0, 100, 200, 400 MeV$, respectively}\\
 \indent b) \footnotesize{$T = 0\div 400$ MeV and $\Phi = -300 \div 300$ MeV.}
\end{center}
 \begin{center}
 \includegraphics[width=7cm, height=4.5cm]{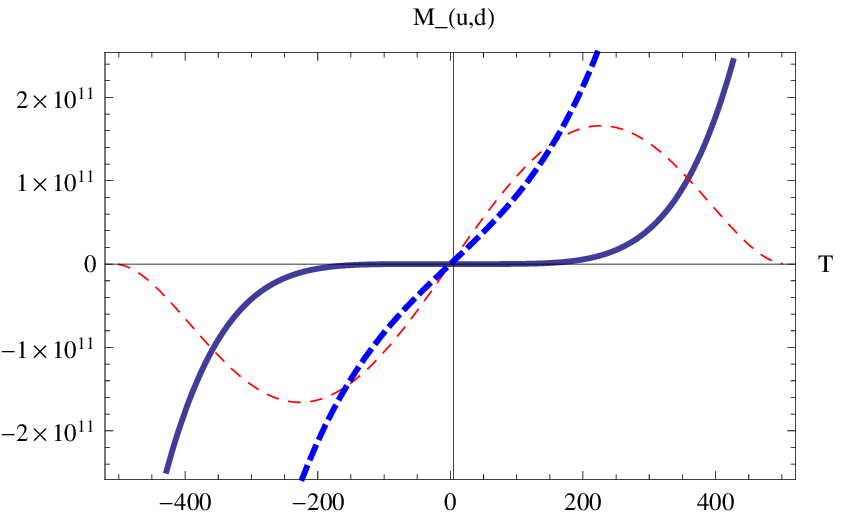}\\
  (a)\\
  \begin{tabular}{cc}
 \includegraphics[width=6.5cm, height=5cm]{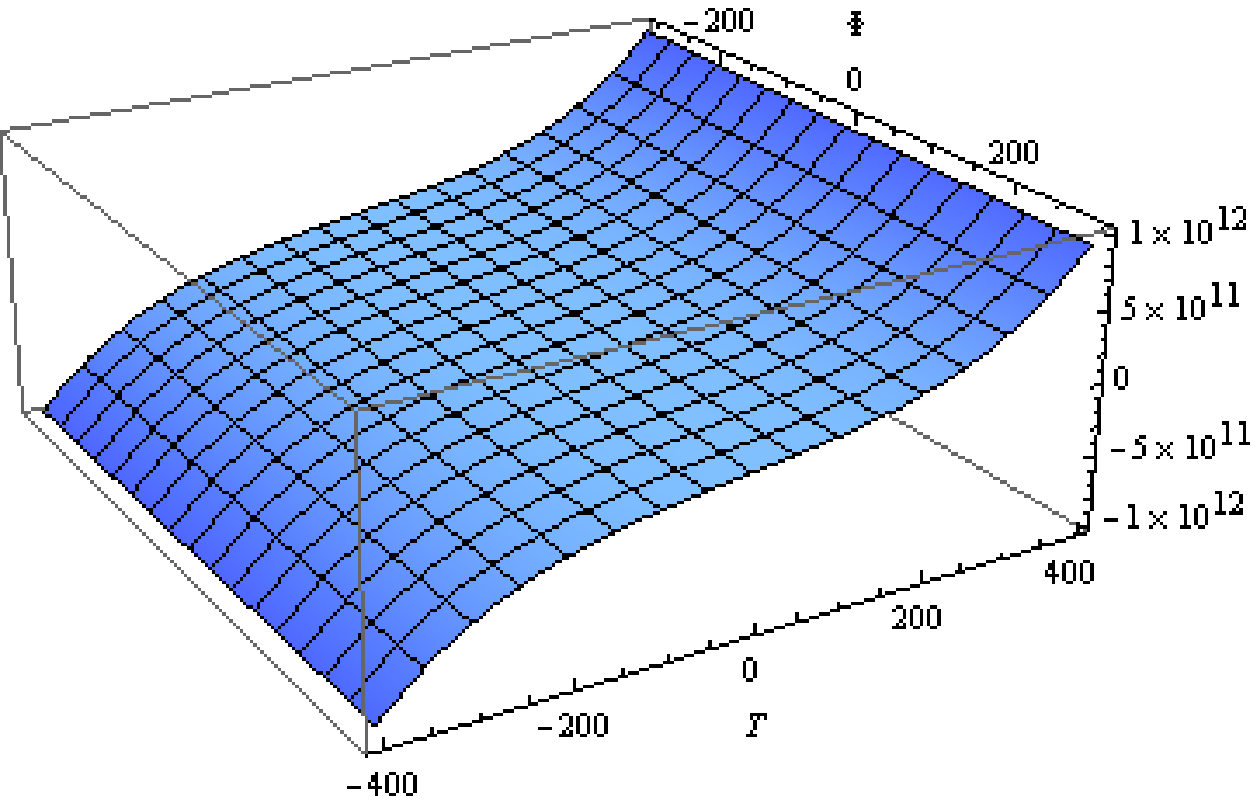}&\includegraphics[width=6.5cm,height=5cm]{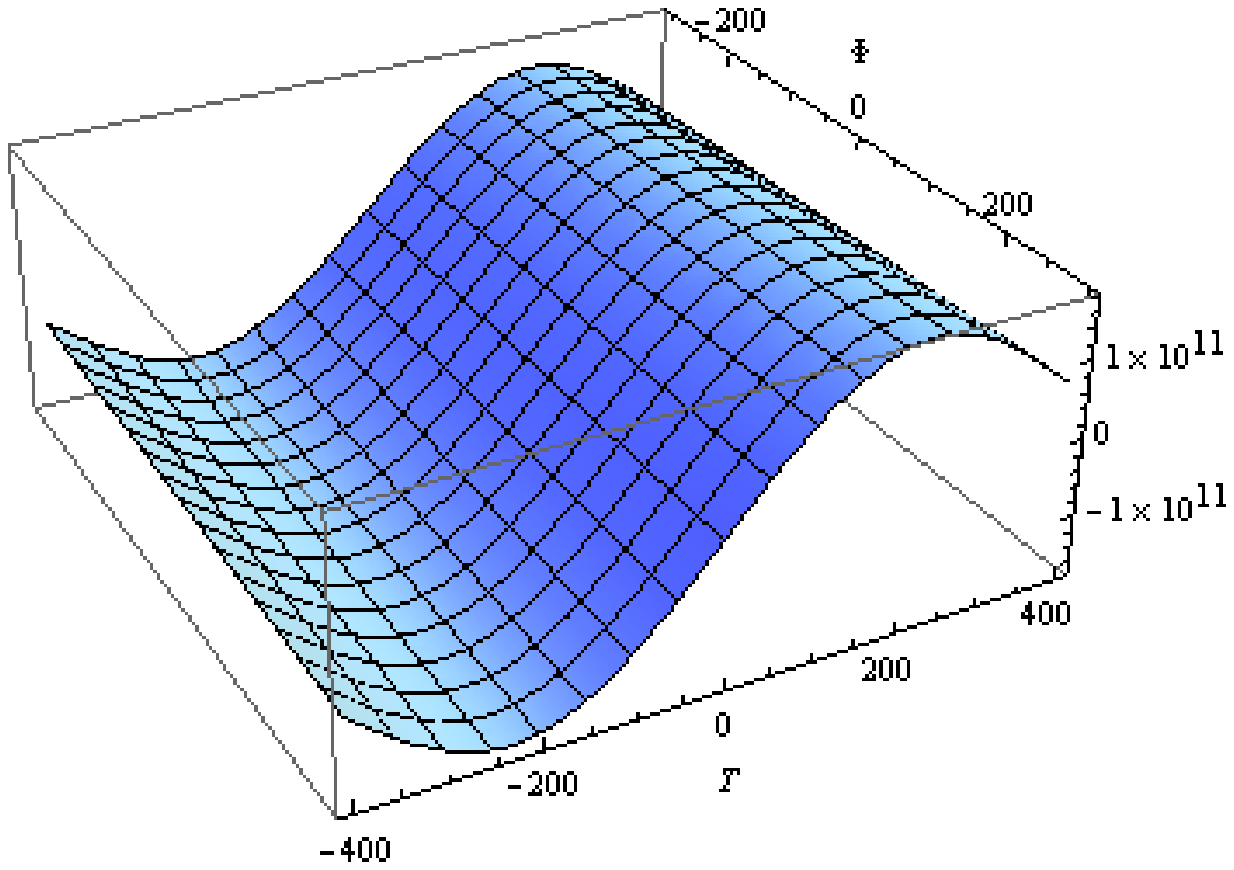}\\
 b) & c)
\end{tabular}\\
\vspace{0.4cm} \footnotesize{\textbf{Fig.3.} The squared scalar mass
$\mathfrak{M}^2$ as a function of temperature T.}
\end{center}
\vspace{-1.5cm}
\begin{flushleft}
\indent a) \footnotesize{$T
= -400 \div 400 MeV$ with $\Phi = 100MeV$ in cases $\mu=m,\mu=1,5m,\mu=0,5m$.}\\
 \indent b,c) \footnotesize{$T = -400\div 400$ MeV and $\Phi = -300 \div 300$
MeV in cases $\mu=1,2m,\mu=0,7m$, respectively.}
\end{flushleft}
\pagebreak
\begin{center}
\begin{tabular}{cc}
 \includegraphics[width=6cm,height=5cm]{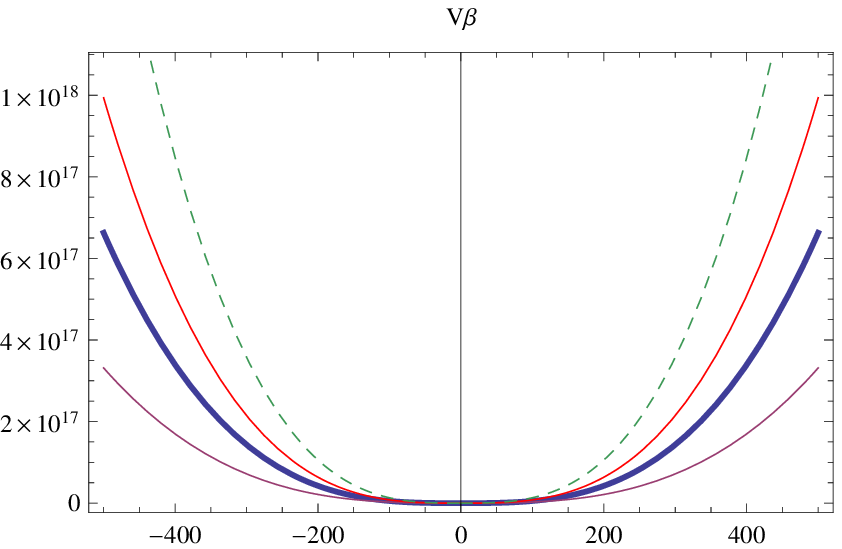}
 &\hspace{1cm}\includegraphics[width=6cm,height=5cm]{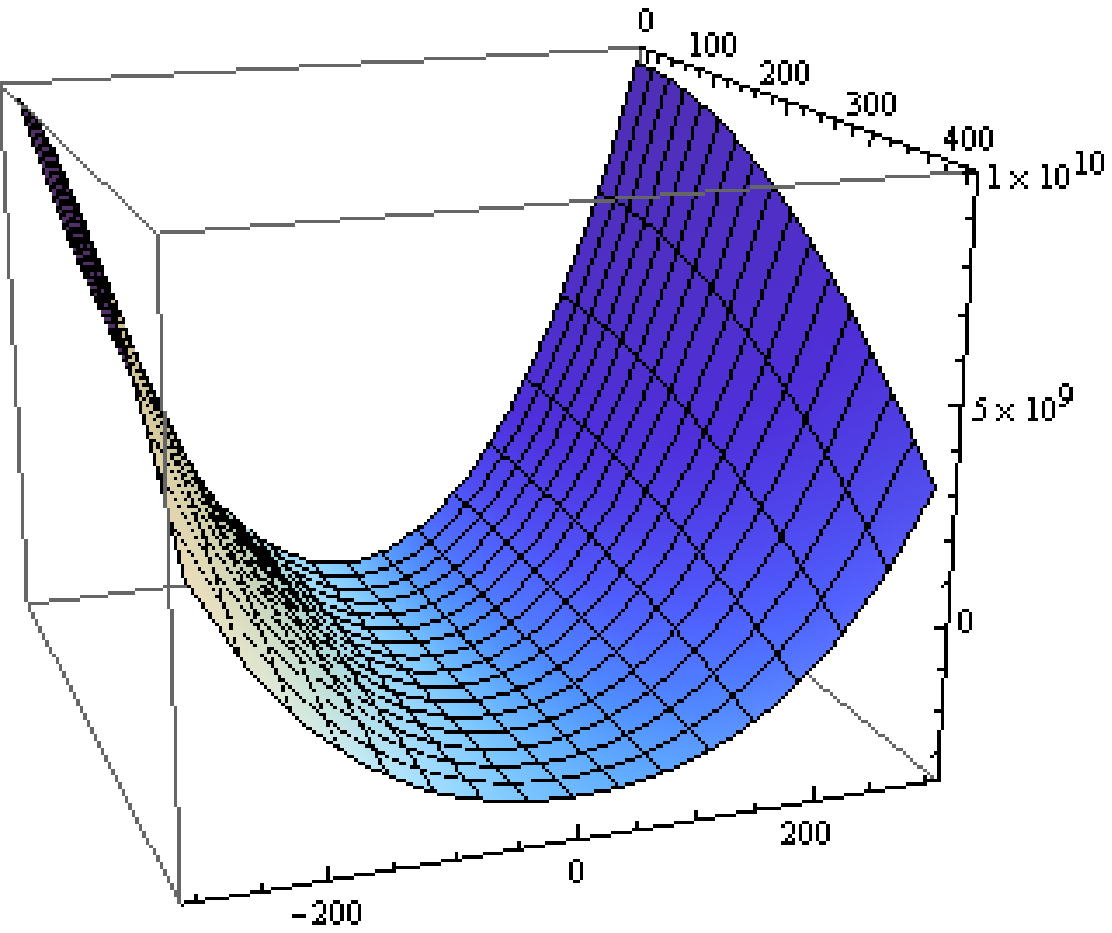}\\
 a)&\hspace{1cm}b)
\end{tabular}\\
\footnotesize{\textbf{Fig.4.} The effective potential $V_\beta$ as a
function of $\mu$ and $\Phi$ at $T =400 MeV$. }\\
\indent \hspace{28pt}a) \footnotesize{$\mu = -500 \div 500 MeV$ with $\Phi = 0, 100, 200, 400 MeV$}, respectively.\\
 \indent \hspace{-50pt}b) \footnotesize{$\mu = 0\div 400$ MeV and $\Phi = -300 \div 300$ MeV.}
\end{center}
\vspace{-3cm}
\begin{center}
\includegraphics[width=8cm,height=7cm]{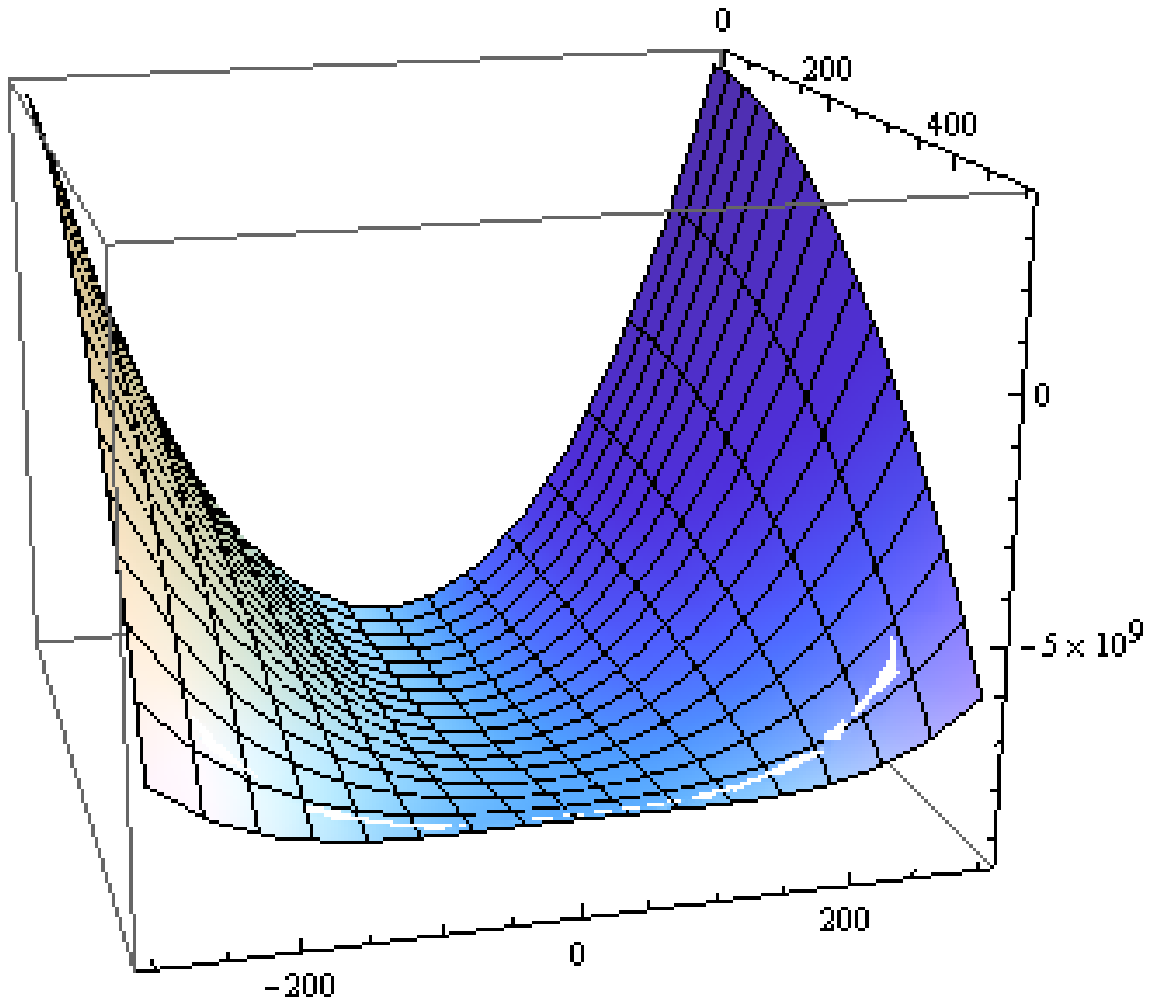}\\
\footnotesize{\textbf{Fig.5.} The effective mass as a function of
chemical potential. It expresses the first order \\phase
transition.The discontinuity is phase transition region (white
region).}
\\\footnotesize{Plot for $\mu = 0\div 500$ MeV and $\Phi = -300 \div 300$
MeV.}
\end{center}
\enlargethispage{4cm} \vspace{-3.5cm}
\begin{center}
\includegraphics[width=14.5cm,height=8.5cm]{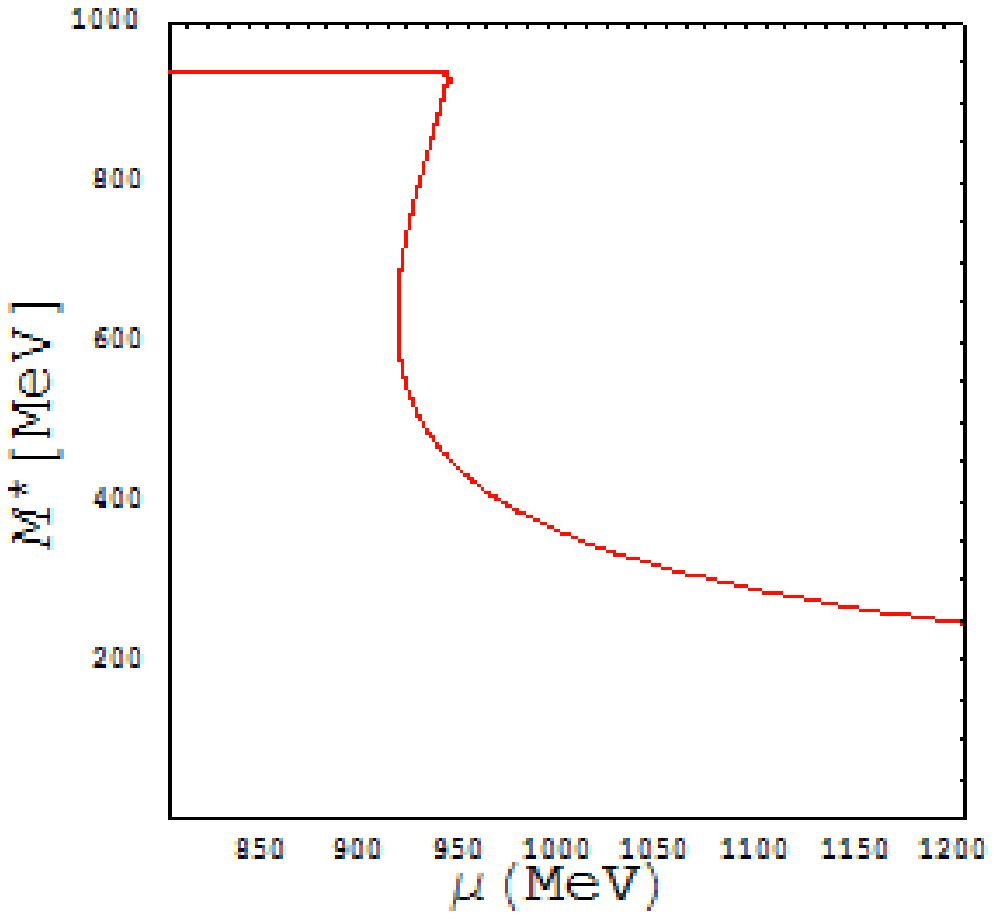}\\
\vspace{-2cm} \footnotesize{\textbf{Fig.6.} Dependence of the
effective squared scalar mass on $\mu$ at fixed
temperature.\\
It expresses the first order phase transition.}
\\\footnotesize{Plot for $\mathfrak{M}^2$ with $\mu
= 0\div 1200$ MeV.}
\end{center}
\pagebreak
\section{DISCUSSION AND CONCLUSION}
We have studied the free energy as a function of temperature and
non-zero chemical potential by background gauge field method in
frame of Abelian theories. Hence the mechanism of cosmological phase
transition is investigated. The graphic solutions have shown that in
the early Universe it is first order phase transition. Furthermore,
after the spontaneous symmetric breaking, one also can see
non-restoration of symmetry in hot gauge field theories for
Cosmology, i.e the Universe is asymmetry.\\
\indent We should mention that one can  study the quantum effects in
phase transition and find exactly the  cosmological critical
temperature if a suitable parameter set is given.
\section*{ACKNOWLEDGMENT}
One of the authors (PHL) would like to thank Prof. Tran Huu Phat for
helpful suggestion  of this problem.

\end{document}